\RequirePackage{fix-cm}
\documentclass{svjour3}                     
\smartqed  
\usepackage{graphicx}
\usepackage{amssymb}
\usepackage{epsfig}
\usepackage{subfigure}
\begin{document}

\title{Area Functional Relation for 5D-Gauss-Bonnet-AdS Black Hole  
}
\author{Parthapratim Pradhan}



\institute{    \at
               Department of Physics\\
               Vivekananda Satavarshiki Mahavidyalaya\\
               (Affiliated to Vidyasagar University)\\
                West Midnapur, West Bengal~721513, India \\
               \email{pppradhan77@gmail.com}
          }

\date{Received: date / Accepted: date}

\maketitle

\begin{abstract}
We present \emph{area (or entropy) functional relation}  for multi-horizons five dimensional (5D) 
Einstein-Maxwell-Gauss-Bonnet-AdS Black Hole. It has been observed by exact and explicit calculation 
that some complicated function of two or three horizons area  is \emph{mass-independent} whereas 
the entropy product relation is \emph{not} mass-independent. 
We also study the local thermodynamic stability of this black hole. The phase transition 
occurs at certain condition. \emph{Smarr mass formula} and \emph{first law} of thermodynamics have been 
derived. This \emph{mass-independent} relation suggests they could turn out to be 
an \emph{universal} quantity and further helps us to understanding the nature of black hole entropy 
(both interior and exterior) at the microscopic level. In the \emph{Appendix}, we have derived the
thermodynamic products for 5D Einstein-Maxwell-Gauss-Bonnet black hole with \emph{vanishing} cosmological 
constant.
\end{abstract}


\maketitle

\section{Introduction}
Black holes are the most intriguing objects in the universe. Thermodynamic products particularly, 
area (or entropy) product of this object is also intriguing when they turned out to be independent 
of the black hole (BH) mass or ADM (Arnowitt-Deser-Misner) mass of the space-time. These are very
popular topic in recent years both in the general relativity community
\cite{ah09,mv13,cdgk,tian,don,pp14,hor15,pp16} and in  the String/M-theory community \cite{cgp11}. 
For example, the product of the inner horizon area (${\cal A}_{2}$) and outer horizon area (${\cal A}_{1}$) 
of Kerr-Newman (KN) BH is\cite{ah09}
\begin{eqnarray}
{\cal A}_{2}{\cal A}_{1} &=& 64 \pi^2 J^2+16\pi^2Q^4 ~.\label{prKN}
\end{eqnarray}
The above relation should indicated that the product does depend on the quantized angular momentum and quantized charges 
of the BH respectively. But it is indeed independent of the BH mass or ADM  mass.  Since only this unique product relation 
is defined on the horizon and is independent of the mass. Therefore, it should be treated  as an ``universal'' quantity. 
More appropriately we can say, a formula only involving horizon 
[inner horizon $({\cal H}^{2})$ and outer horizon $({\cal H}^{1})$] quantities \emph{could} turn out 
to be \emph{universal} and this is only a \emph{necessary condition}. 

In the case of Eq. \ref{prKN} it has actually been \emph{shown} that the same equation also holds for much more general 
space-times than the KN electrovacuum solution. But this does not follow from the observation that the formula is 
mass-independent alone. The mass-independence should by no means be expected to be 
sufficient. In particular, for the mass-independence relation that we will be derived in this work, it may be that they turn 
out to be ``universal'', i.e. they could be valid for more general solutions, but this would require a rigorous proof. So far, 
they only have a chance to be ``universal'', since they satisfy the necessary condition.

For extremal KN BH (where both horizons area coincide i.e. ${\cal A}_{1}={\cal A}_{2}$), these relations become mass-independent 
inequality:
\begin{eqnarray}
{\cal A}_{1} &\geq & \sqrt{64 \pi^2 J^2+16\pi^2Q^4} ~.\label{KN1}
\end{eqnarray}
However, Eq. \ref{KN1} is valid for \emph{all} KN BHs (considering only ``true'' BHs that really have an event horizon, i.e. 
no over extremal objects). In the special case of an extremal BH, one has equality in Eq. \ref{KN1}.

When we have taken into account the BPS (Bogomol'nyi-Prasad-Sommerfield) states 
\footnote{These states are supersymmetric states where mass equal to the central charge $Q$. They are saturated by BPS bound. 
If the supersymmetry is not broken then $M=|Q|$ otherwise $M>|Q|$. These states are stable in extremal BH physics. These states 
playing also a key role in QFT and string theory.}, the area product should be \cite{cgp11}
\begin{eqnarray}
{\cal A}_{1} {\cal A}_{2}  &=& \left(8\pi {\ell _{pl}}^2\right)^2\left(\sqrt{N_{1}}+\sqrt{N_{2}}\right)
\left(\sqrt{N_{1}}-\sqrt{N_{2}}\right)= 
N , \,\, N\in {\mathbb{N}}, N_{1}\in {\mathbb{N}}, N_{2} \in {\mathbb{N}} ~.\label{ppl}
\end{eqnarray}
where the integers $N_{1}$ and $N_{2}$ are described as excitation numbers of the left and 
right moving sectors of a weakly-coupled two-dimensional CFT (conformal field theory). 

Recently, Page et al. \cite{don} suggested  a heuristic argument for the universal area products
of a four-dimensional adiabatically distorted KN BH. They showed for an adiabatically distorted 
multi-horizon BH, the product of horizon areas could be expressed in terms of a polynomial function 
of charges, angular mometa and inverse square root of the cosmological constant.   It has been argued in 
\cite{cgp11} if the cosmological parameter is quantized, the product of ${\cal H}^{1}$ area and ${\cal H}^{2}$ 
area could provide a ``looking glass for probing the microscopics of general BHs''.

Now by introducing the cosmological constant, we have examined in this work whether the conjectured 
``area product is universal'' does hold for 5D cosmological Einstein-Maxwell-Gauss-Bonnet (EMGB) 
BH which is described in an alternative theory of gravity. Among the many higher-dimensional generalisations 
of general theory of relativity, Gauss-Bonnet theory is probably a more promising candidate, since the field 
equations are still second-order equations, despite the appearance of additional curvature terms in the underlying 
action. Thus this theory is more important than among other dubious theories like $f(R)$ theories.

It has already been shown by explicitly in \cite{mv13} for Reissner Nordstr\"{o}m-AdS BH, 
the area product is not mass-independent (universal). But slightly complicated function of 
horizon area i.e $f({\cal A}_{1},{\cal A}_{2})$ is indeed independent of mass. It has been 
explicitly calculated in\cite{mv13} for RN-dS BH, the functional form should read:
\begin{eqnarray}
f({\cal A}_{1},{\cal A}_{2},{\cal A}_{3})&=& 
\frac{(\sqrt{{\cal A}_{1}}+\sqrt{{\cal A}_{2}}+\sqrt{{\cal A}_{3}})\sqrt{{\cal A}_{1}}
\sqrt{{\cal A}_{2}}\sqrt{{\cal A}_{3}}}{{\cal A}_{1}+{\cal A}_{2}+{\cal A}_{3}+
\sqrt{{\cal A}_{1}{\cal A}_{2}}+\sqrt{{\cal A}_{2}{\cal A}_{3}}+\sqrt{{\cal A}_{3} {\cal A}_{1}}} 
=4\pi Q^2 ~.\label{v1}
\end{eqnarray}
\footnote{It seems to be a typo in Eq. (64) in \cite{mv13}. The term ${\cal A}_{1}^2+{\cal A}_{2}^2+{\cal A}_{3}^2+
\sqrt{{\cal A}_{1}{\cal A}_{2}}+\sqrt{{\cal A}_{2}{\cal A}_{3}}+\sqrt{{\cal A}_{3} {\cal A}_{1}}$ should read 
there ${\cal A}_{1}+{\cal A}_{2}+{\cal A}_{3}+\sqrt{{\cal A}_{1}{\cal A}_{2}}+\sqrt{{\cal A}_{2}{\cal A}_{3}}
+\sqrt{{\cal A}_{3} {\cal A}_{1}}$.}
where ${\cal A}_{3}$ is the area of the cosmological horizon.

For RN-AdS BH,  the functional form should be\cite{mv13}:
\begin{eqnarray}
f({\cal A}_{1},{\cal A}_{2})&=& \left[1+\frac{|\Lambda|}{12\pi}
\left({\cal A}_{1}+{\cal A}_{2}+\sqrt{{\cal A}_{1}{\cal A}_{2}}\right)\right]\sqrt{{\cal A}_{1}{\cal A}_{2}}
=4\pi Q^2 ~.\label{v2}
\end{eqnarray}
\footnote{Again in Eq. (86), there is a typo. The term $(\sqrt{{\cal A}_{1}}+\sqrt{{\cal A}_{2}})^2$ should read there
$\left({\cal A}_{1}+{\cal A}_{2}+\sqrt{{\cal A}_{1}{\cal A}_{2}}\right)$.}

It has been elaborated by Hennig for KN-AdS BH\cite{jh}, this functional form should be
$$
f({\cal A}_{1},{\cal A}_{2})= \left[1+\frac{|\Lambda|}{6\pi}({\cal A}_{1}+{\cal A}_{2}+4\pi Q^2)+
\frac{|\Lambda|^2}{144\pi^2}({\cal A}_{1}^2+{\cal A}_{1}{\cal A}_{2}+{\cal A}_{2}^2)\right]
{\cal A}_{1}{\cal A}_{2} \nonumber\\
$$
\begin{eqnarray}
 &=& 64\pi^2J^2+16\pi^2Q^4 ~.\label{vv}
\end{eqnarray}
Interestingly, the above three functional form is independent of mass.

In this work we would like to investigate  what is the form of this function for 5D EMGB-AdS BH? 
It should be noted that in the Einstein gravity, the famous Bekenstein-Hawking \cite{bk73,bcw73} 
relation is satisfied by the equation ${\cal S}=\frac{\cal A}{4}$. Therefore the area product 
is simply connected to the entropy product. Whereas in the Gauss-Bonnet 
gravity the area product simply is not proportional to the entropy product of the 
${\cal H}^{2}$ and ${\cal H}^{1}$. We would like to discuss this issue in the Sec. 2. 

There are some more complicated function of $f({\cal A}_{1},{\cal A}_{2})$ that are mass independent, but 
generically they are not simple function of ${\cal H}^{1}$ area and ${\cal H}^{2}$ area 
(whereas for Kerr BH and KN BH a simple area product is sufficient). 
It should be mentioned that mass independent function of physical horizon areas in spherically symmetry 
space-time is connected to the quasi-local mass $m(r)$ of any generic Laurent polynomial. 
Again this quasi-local mass is being related to the Laurent polynomial \cite{mv13} of the aerial 
radius $r_{i}=\sqrt{\frac{{\cal A}_{i}}{4\pi}}=\sqrt{\frac{{\cal S}_{i}}{\pi}}$. 

The Laurent polynomial \footnote{A Laurent polynomial is a linear algebric polynomial in a field ${\mathbb F}$ with 
positive and negative power combination of a variable in ${\mathbb F}$ i.e. 
$p=\sum_{i=-n}^{i=+n} a_{i}x^{i}$ 
and $a_{i} \in {\mathbb F}$. } is proportional to the normalized symmetric 
elementary polynomial so that its highest 
degree coefficient is unity and its lowest degree coefficient is a constant $c$ where $c\neq 0$. If there are $N$ 
horizons then the quasilocal mass is a Laurent polynomial from which one may obtain $N-1$ mass-independence quantities 
in terms of the horizon radius. Now if $d$ of these horizons are virtual and if that can be eliminated by any way
then one can find $N-d-1$ mass independent quantities and it can be expressed in terms of only two physical horizons radii 
(although it is some complicated combination of horizon area).

In the Appendix, we have provided thermodynamic product relations of EMGB-AdS BH with $\Lambda=0$. Particularly, we have 
derived the specific heat, Komar energy, Smarr mass formula, first law of thermodynamics and Smarr-Gibbs-Duhem relation 
of ${\cal H}^{2}$ and ${\cal H}^{1}$ that has not been studied before and also not been discussed in \cite{ar}.

The organization of the paper is as follows. In the next section, we provide the thermodynamic 
properties of five dimensional EMGB-AdS BH. In Sec. 3, we described
the entropy functional relation in 5D EMGB-AdS Space-time. In Sec. 4, we have given the conclusions. 
Finally in the \emph{Appendix section} (Sec. 5), we have studied the thermodynamic properties of EMGB-AdS BH 
in the limiting case of $\Lambda=0$.

\section{5D Einstein-Maxwell-Gauss-Bonnet-AdS BH:}
The action for the five dimensional (5D) Einstein-Maxwell-Gauss-Bonnet-AdS BH is given by 
\begin{eqnarray}
I &=& \frac{1}{16\pi G_{5}}\int d^5x\sqrt{-g}\left[R 
  + \alpha (R_{abcd}R^{abcd} -4 R_{ab}R^{ab}+R^2)+L_{matter}\right]  ~.\label{eq}
\end{eqnarray}
where $G_{5}$ is the Newtonian constant in five dimension, $\alpha$ is the Gauss-Bonnet 
coupling constant, $L_{matter}=F_{ab}F^{ab}-2\Lambda$, $F_{ab}$ denotes the 
electromagnetic Faraday tensor and $\Lambda$ is the Cosmological constant.  The Gauss-Bonnet 
coupling constant has many interesting features. For example, in higher curvature gravity the 
equation of motion consists only the second order derivatives of the metric. But in four dimensions, this 
coupling constant $\alpha$ is present in the metric function but surprisingly it is  absent in horizon radii 
and therefore it is interesting in the sense that it is a non-trivial 
consequences of the generalized Gauss-Bonnet theory. It may be noted  that this $\alpha$ term is naturally 
comes from the heterotic superstring theory \cite{zw} and plays a crucial role in Cherns-Simons theory.

A five dimensional version of spherically symmetric solution in EMGB gravity with the 
Cosmological constant is represented by the metric\cite{deser,wilt,wheeler,cai,dehmami,taz}
\begin{eqnarray}
ds^2=-{\cal F}(r)dt^{2}+\frac{dr^{2}}{{\cal F}(r)}+r^{2}d\Omega_{3}^2 
~.\label{gb}
\end{eqnarray}
where $d\Omega_{3}^2$ denotes the line elements of a 3D hypersurface with constant curvature 
$6k$ and $k=-1,0,+1$ represents the hyperbolic, planar and spherical topology of the BH 
horizons respectively. 
The above metric function ${\cal F}(r)$ is defined by
\begin{eqnarray}
{\cal F}(r) &=& k+\frac{r^2}{4\alpha}-\frac{r^2}{4\alpha}
\sqrt{1+\frac{8\alpha M}{r^4}-\frac{8\alpha Q^2}{3r^6}+\frac{4\alpha \Lambda}{3}} 
~.\label{gb1}
\end{eqnarray}
Here $M$ and $Q$ denotes the mass and charge of the BH respectively. We are restricted 
here in the case of spherical topology i.e. $k=1$. The BH horizons correspond to
${\cal F}(r)=0$ i.e. 
\begin{eqnarray}
\frac{\Lambda}{3}r^6-2r^4+2(M-2\alpha)r^2-\frac{2}{3}Q^2 &=& 0
~.\label{gb2}
\end{eqnarray}
For convenient, let us put $r^2=x$ and $\Lambda=\frac{1}{\ell^2}$. Then the above Eq. 
reduces to 
\begin{eqnarray}
x^3-6 \ell^2 x^2+6(M-2\alpha)\ell^2 x-2 \ell^2 Q^2 &=& 0 ~.\label{gb3}
\end{eqnarray}
The positions of the BH horizons are related to zeros of the above polynomial and they are the square roots 
of these zeros. Hence only real and positive zeros are relevant in order to obtain ``physical'' horizons and not 
just ``virtual'' horizons described by complex coordinate positions, as sometimes described in the literature \cite{mv13}.
Hence, before deriving relations for the horizons, we should first discuss the nature of these zeros depending the BH 
parameter values. Since the polynomial is negative at $x=0$ and diverges to $+\infty$ for $x \rightarrow \infty$ there 
is certainly always at least one positive real zero $x$.  

\emph{Case I:}

For the BH parameter values $\Lambda=\frac{1}{\ell^2}>0$ and $M>2\alpha$ and using Mathematica we find the exact 
only one real positive root of the Eq. \ref{gb3} is
\begin{eqnarray}
x_{1} &=& 2\ell^2+\left[8\ell^6-6(M-2\alpha)\ell^4+\ell^2Q^2+z\right]^\frac{1}{3}+
\frac{2\ell^2\left(2\ell^2-M+2\alpha \right)}{\left[8\ell^6-6(M-2\alpha)\ell^4+\ell^2Q^2+z\right]^\frac{1}{3}} \nonumber\\
~.\label{gb5}
\end{eqnarray}
where 
\begin{eqnarray}
z &=& \left[8\ell^6(M-2\alpha)^3-12\ell^8(M-2\alpha)^2-12\ell^6(M-2\alpha)Q^2+16\ell^8Q^2+\ell^4Q^4\right]^\frac{1}{2} \nonumber\\
\end{eqnarray}
In terms of area ${\cal A}_{i}=2\pi^2r_{i}^3, (i=1,2,3)$ and cosmological constant, the equation 
may be rewritten as 
\begin{eqnarray}
f({\cal A}_{1}) &=& \left(\frac{{\cal A}_{1}}{2\pi^2}\right)^{\frac{2}{3}} \nonumber\\
 &=& \frac{2}{\Lambda}+
\left[\frac{8}{\Lambda^3}-\frac{6(M-2\alpha)}{\Lambda^2}+\frac{Q^2}{\Lambda}+z\right]^\frac{1}{3}
+\frac{\frac{2}{\Lambda}\left(\frac{2}{\Lambda}-M+2\alpha \right)}
{\left[\frac{8}{\Lambda^3}-\frac{6(M-2\alpha)}{\Lambda^2}+\frac{Q^2}{\Lambda}+z\right]^\frac{1}{3}} \nonumber\\
 ~.\label{gb6}
\end{eqnarray}
where 
\begin{eqnarray}
z &=& \left[\frac{8(M-2\alpha)^3}{\Lambda^3}-\frac{12(M-2\alpha)^2}{\Lambda^4}-\frac{12(M-2\alpha)Q^2}{\Lambda^3}
+\frac{16Q^2}{\Lambda^4}+\frac{Q^4}{\Lambda^2}\right]^\frac{1}{2} \nonumber\\
\end{eqnarray}
Unfortunately, this equation is explicitly \emph{mass dependent}. 

\emph{Case II:}

If we take the parameter values $\Lambda=-\frac{1}{\ell^2}$ and $M>2\alpha$ then the Eq. \ref{gb3} becomes 
\begin{eqnarray}
x^3+6 \ell^2 x^2-6(M-2\alpha)\ell^2 x+2 \ell^2 Q^2 &=& 0 ~.\label{bb}
\end{eqnarray}
Now using Vieta's theorem, we find the following three Eqs.:
\begin{eqnarray}
x_{1}+x_{2}+x_{3} &=& -6\ell^2 ~.\label{eq1}\\
x_{1}x_{2}+x_{1}x_{3}+x_{2}x_{3} &=& -6(M-2\alpha)\ell^2 ~.\label{eq2}\\
x_{1}x_{2}x_{3} &=& -2\ell^2Q^2 ~.\label{eq3}
\end{eqnarray}
If there exist two physical horizons and if we eliminate the third root 
from (\ref{eq1}) and (\ref{eq3}), we get the following relation in terms of 
two horizon:
\begin{eqnarray}
x_{1}x_{2}(x_{1}+x_{2}) &=& 2\ell^2\left(Q^2-3x_{1}x_{2}\right) ~.\label{eq5}
\end{eqnarray}
In terms of two physical horizons area, the \emph{mass independent}, coupling constant independent but 
cosmological constant dependent and charge dependent relation is 
\begin{eqnarray} 
f({\cal A}_{1},{\cal A}_{2}) &=& \frac{
\left(\frac{{\cal A}_{1}{\cal A}_{2}}{4\pi^4}\right)^{\frac{2}{3}}
\left[\left(\frac{{\cal A}_{1}}{2\pi^2}\right)^{\frac{2}{3}}+
\left(\frac{{\cal A}_{2}}{2\pi^2}\right)^{\frac{2}{3}}\right]}
{\left[Q^2-3\left(\frac{{\cal A}_{1}{\cal A}_{2}}{4\pi^4}\right)^{\frac{2}{3}}\right] } 
=\frac{2}{\Lambda} ~.\label{v4}
\end{eqnarray}
From (\ref{eq1}) and (\ref{eq2}), in terms of two horizons, \emph{mass dependent}, Cosmological 
constant dependent and coupling constant dependent relation is 
\begin{eqnarray}
x_{1}x_{2}-\frac{6}{\Lambda}(x_{1}+x_{2}) -(x_{1}+x_{2})^2 &=& \frac{12\alpha}{\Lambda}-\frac{6M}{\Lambda}
{\Lambda} ~.\label{eq6}
\end{eqnarray}
In terms of two horizons area, this should be re-written as
$$
\left(\frac{{\cal A}_{1}{\cal A}_{2}}{4\pi^4}\right)^{\frac{2}{3}}-\frac{6}{\Lambda}
\left[\left(\frac{{\cal A}_{1}}{2\pi^2}\right)^{\frac{2}{3}}+
\left(\frac{{\cal A}_{2}}{2\pi^2}\right)^{\frac{2}{3}}\right]-
$$
\begin{eqnarray} 
\left[\left(\frac{{\cal A}_{1}}{2\pi^2}\right)^{\frac{2}{3}}+
\left(\frac{{\cal A}_{2}}{2\pi^2}\right)^{\frac{2}{3}}\right]^2
&=& \frac{12\alpha}{\Lambda}-\frac{6M}{\Lambda} ~.\label{v6}
\end{eqnarray}

It should be noted that $\Lambda, M,\alpha, Q$ dependent relation is
\begin{eqnarray}
x_{1}x_{2}-\frac{2Q^2}{\Lambda}\left(\frac{x_{1}+x_{2}}{x_{1}x_{2}}\right) &=& 
\frac{12\alpha}{\Lambda}-\frac{6M}{\Lambda} ~.\label{eq7}
\end{eqnarray}
If we working on area, this can be written as
\begin{eqnarray} 
\left(\frac{{\cal A}_{1}{\cal A}_{2}}{4\pi^4}\right)^{\frac{2}{3}}-
\frac{2Q^2}{\Lambda}\left[\left(\frac{{\cal A}_{1}}{2\pi^2}\right)^{\frac{2}{3}}+
\left(\frac{{\cal A}_{2}}{2\pi^2}\right)^{\frac{2}{3}}\right] 
&=& \frac{12\alpha}{\Lambda}-\frac{6M}{\Lambda} ~.\label{v7}
\end{eqnarray}

\emph{Case III:}

If we take the parameter values $\Lambda=-\frac{1}{\ell^2}$ and $M<2\alpha$ then the Eq. \ref{gb3} becomes 
\begin{eqnarray}
x^3+6 \ell^2 x^2+6(M-2\alpha)\ell^2 x+2 \ell^2 Q^2 &=& 0 ~.\label{cc}
\end{eqnarray}
Again using Vieta's theorem, one obtains
\begin{eqnarray}
x_{1}+x_{2}+x_{3} &=& -6\ell^2 ~.\label{ec1}\\
x_{1}x_{2}+x_{1}x_{3}+x_{2}x_{3} &=& 6(M-2\alpha)\ell^2 ~.\label{ec2}\\
x_{1}x_{2}x_{3} &=& -2\ell^2Q^2 ~.\label{ec3}
\end{eqnarray}
Eliminating third root and if there exists two physical horizons one obtains mass-independent relation as 
we find previously in Eq. \ref{eq5} and Eq. \ref{v4}.

If there exists three horizons then eliminating $\ell$ from (\ref{ec1}) and (\ref{ec3}), we find 
the following relation:
\begin{eqnarray}
\frac{x_{1}+x_{2}+x_{3}}{x_{1}x_{2}x_{3}} &=& \frac{3}{Q^2} ~.\label{eq4}
\end{eqnarray}
In terms of area the mass independent functional form is
\begin{eqnarray}
f({\cal A}_{1},{\cal A}_{2},{\cal A}_{3}) &=& 
\frac{3\left(\frac{{\cal A}_{1}}{2\pi^2}\right)^{\frac{2}{3}} 
\left(\frac{{\cal A}_{2}}{2\pi^2}\right)^{\frac{2}{3}} 
\left(\frac{{\cal A}_{3}}{2\pi^2}\right)^{\frac{2}{3}}}
{\left(\frac{{\cal A}_{1}}{2\pi^2}\right)^{\frac{2}{3}}
+\left(\frac{{\cal A}_{2}}{2\pi^2}\right)^{\frac{2}{3}}+ 
\left(\frac{{\cal A}_{3}}{2\pi^2}\right)^{\frac{2}{3}}}
= Q^2 ~.\label{v3}
\end{eqnarray}
This is explicitly \emph{mass independent}, coupling constant independent 
and cosmological constant independent but charge dependent relation. 
But it is a complicated function of three horizons area. 

Finally, the other \emph{mass independent} relation in terms of three horizons area are 
\begin{eqnarray}
\left(\frac{{\cal A}_{1}}{2\pi^2}\right)^{\frac{2}{3}}
+\left(\frac{{\cal A}_{2}}{2\pi^2}\right)^{\frac{2}{3}}+ 
\left(\frac{{\cal A}_{3}}{2\pi^2}\right)^{\frac{2}{3}} 
 &=& - \frac{6}{\Lambda} ~.\label{v8}
\end{eqnarray}
and 
\begin{eqnarray}
\left({\cal A}_{1}{\cal A}_{2}{\cal A}_{3}\right)^{\frac{2}{3}} &=& -
\frac{8\pi^4Q^2}{\Lambda} ~.\label{v9}
\end{eqnarray}

\section{Entropy Functional relation in 5D EMGB-AdS Space-time:}
Since we have already mentioned in the introduction that in the higher curvature gravity
the Bekenstein-Hawking relation do not work because the entropy relation get modified by 
the following relation:
\begin{eqnarray}
{\cal S}_{i} &=& \frac{\Omega_{d-2} r_{i}^{d-2}}{4}
\left(1+\frac{2(d-2)(d-3)\alpha k} {r_{i}^2}\right)
~.\label{v10}
\end{eqnarray}
where $\Omega_{d-2}=\frac{2\pi^{\frac{(d-1)}{2}}}{\Gamma(\frac{d-1}{2})}$ and the 
area of the horizon becomes ${\cal A}_{i}=\Omega_{d-2}r_{i}^{d-2}$.
For 5D EMGB-AdS Space-time the modified entropy relation is
\begin{eqnarray}
{\cal S}_{i} &=& \frac{2\pi^2 r_{i}^3}{4}\left(1+\frac{12\alpha k} {r_{i}^2}\right)
~.\label{v11}
\end{eqnarray}

In terms of area this should be written as
\begin{eqnarray}
{\cal S}_{i}  &=& \frac{{\cal A}_{i}}{4}+
6\pi^2 \alpha k\left(\frac{{\cal A}_{i}}{2\pi^2}\right)^{\frac{1}{3}} ~.\label{v12}
\end{eqnarray}
When $\alpha$ goes to zero, we get the Bekenstein-Hawking entropy relation as usual.
Now we compute the product of the entropy in 5D EMGB-AdS BH and it should be 
\begin{eqnarray}
{\cal S}_{1}{\cal S}_{2}{\cal S}_{3} &=& \frac{\pi^6}{2\sqrt{2}}\frac{Q}{\Lambda^{\frac{3}{2}}}
\left[Q^2+36\alpha k(M-2\alpha)+432 k^2\alpha^2+864k^3\alpha^3 \Lambda \right]
~.\label{v13}
\end{eqnarray}
It suggests that the product is strictly depends upon the mass parameter hence the entropy 
product is not universal in 5D EMGB-AdS BH. This is a curious result. 

The entropy sum is given by 
\begin{eqnarray}
{\cal S}_{1}+{\cal S}_{2}+{\cal S}_{3}  &=& \frac{{\cal A}_{1}+{\cal A}_{2}+{\cal A}_{3}}{4}+
6\pi^2 \alpha k \left[\left(\frac{{\cal A}_{1}}{2\pi^2}\right)^{\frac{1}{3}}+
\left(\frac{{\cal A}_{2}}{2\pi^2}\right)^{\frac{1}{3}}+
\left(\frac{{\cal A}_{3}}{2\pi^2}\right)^{\frac{1}{3}}\right]~.\label{v14}
\end{eqnarray}
and other relation is 
$$
{\cal S}_{1} {\cal S}_{2}+{\cal S}_{2} {\cal S}_{3}+{\cal S}_{3} {\cal S}_{1}
= \frac{{\cal A}_{1}{\cal A}_{2}+{\cal A}_{2}{\cal A}_{3}+{\cal A}_{3}{\cal A}_{1}}{16}+
$$
$$
(6\pi^2 \alpha k)^2 \left[\left(\frac{{\cal A}_{1}{\cal A}_{2}}{4\pi^4}\right)^{\frac{1}{3}}+
\left(\frac{{\cal A}_{2}{\cal A}_{3}}{4\pi^4}\right)^{\frac{1}{3}}+
\left(\frac{{\cal A}_{3}{\cal A}_{1}}{4\pi^4}\right)^{\frac{1}{3}}\right] + 
$$
\begin{eqnarray}
\frac{3\pi^2 \alpha k}{2} 
\left[{\cal A}_{1} \left\{\left(\frac{{\cal A}_{2}}{2\pi^2}\right)^{\frac{1}{3}} 
+\left(\frac{{\cal A}_{3}}{2\pi^2}\right)^{\frac{1}{3}} \right \}
+{\cal A}_{2}  \left\{ \left(\frac{{\cal A}_{1}}{2\pi^2}\right)^{\frac{1}{3}}
+\left(\frac{{\cal A}_{3}}{2\pi^2}\right)^{\frac{1}{3}} \right \}
+{\cal A}_{3} \left\{ \left(\frac{{\cal A}_{2}}{2\pi^2}\right)^{\frac{1}{3}}
+\left(\frac{{\cal A}_{1}}{2\pi^2}\right)^{\frac{1}{3}} \right \}
\right] 
~.\label{v15}
\end{eqnarray}
From this entropy functional relation it is quite clear that these formulae interestingly 
\emph{mass independent}.

\subsection{First Law and Specific heat for EMGB-AdS:}
From Eq. (\ref{gb2}), we find the mass parameter in terms of the area of the BH:
\begin{eqnarray}
M &=& 2\alpha +\left(\frac{{\cal A}_{i}}{2\pi^2}\right)^{\frac{2}{3}}+
\frac{Q^2}{3} \left(\frac{2\pi^2}{{\cal A}_{i}}\right)^{\frac{2}{3}}-
\frac{\Lambda}{6}\left(\frac{{\cal A}_{i}}{2\pi^2}\right)^{\frac{4}{3}}
~.\label{v16}
\end{eqnarray}
Therefore the mass differential becomes
\begin{eqnarray}
dM &=& \Upsilon_{i} d{\cal A}_{i}  +\Phi_{i} dQ+ \lambda_{i} d\Lambda
~. \label{v17}
\end{eqnarray}
where
\begin{eqnarray}
\Upsilon_{i} &=& \frac{\partial M}{\partial {\cal A}_{i}}= \frac{2}{3(2\pi^2)^{2/3}} 
\frac{1}{{\cal A}_{i}^{1/3}}\left[1-\frac{(2\pi^2)^{4/3}}{3}\frac{Q^2}{{\cal A}_{i}^{4/3}} 
-\frac{\Lambda}{3(2\pi^2)^{2/3}}{\cal A}_{i}^{2/3} \right]\nonumber \\
\Phi_{i} &=& \frac{\partial M}{\partial Q}=\frac{2Q}{3}\left(\frac{2\pi^2}{{\cal A}_{i}}\right)^{\frac{2}{3}}
\nonumber\\
\lambda_{i} &=& \frac{\partial M}{\partial \Lambda}=-\frac{1}{6}
\left(\frac{{\cal A}_{i}}{2\pi^2}\right)^{\frac{4}{3}}
~. \label{inv}
\end{eqnarray}
and 
where
\begin{eqnarray}
\Upsilon_{i} &=& \mbox{Effective surface tension for the horizons}
\nonumber \\
\Phi_{i} &=& \mbox{Electromagnetic potentials for the horizons }
\nonumber\\
\lambda_{i} &=&  \mbox{Cosmological potentials for the horizons }
~. \label{v18}
\end{eqnarray}
Therefore the first law of thermodynamics  becomes
\begin{eqnarray}
dM &=& \frac{T_{i}}{4} d{\cal A}_{i}  +\Phi_{i} dQ+ \lambda_{i} d\Lambda ~. \label{v19}
\end{eqnarray}
where the BH temperature defined as
\begin{eqnarray}
T_{i} &=& \frac{8}{3(2\pi^2)^{2/3}} 
\frac{1}{{\cal A}_{i}^{1/3}}\left[1-\frac{(2\pi^2)^{4/3}}{3}\frac{Q^2}{{\cal A}_{i}^{4/3}} 
-\frac{\Lambda}{3(2\pi^2)^{2/3}}{\cal A}_{i}^{2/3} \right]~. \label{v20}
\end{eqnarray}

An important quantity in BH thermodynamics is the specific heat which measures the stability of the 
BH. It is defined by 
\begin{eqnarray}
C_{i} &=& \frac{\partial{\cal M}}{\partial T_{i}} .~\label{c1}
\end{eqnarray}
which is found to be for EMGB-AdS BH:
\begin{eqnarray}
C_{i} &=& -\frac{3\pi^2}{2} r_{i}^3 \frac{\left(1-\frac{1}{3}\frac{Q^2}{r_{i}^4}-\frac{\Lambda}{3}r_{i}^2\right)}
{\left(1-\frac{5}{3}\frac{Q^2}{r_{i}^4}+\frac{\Lambda}{3}r_{i}^2\right)}  .~\label{c2}
\end{eqnarray}
Let us explain the above expression of specific heat for a different regime in the parameter space.

Case I: The specific  heat $C_{i}$ is positive when $\left(1-\frac{1}{3}\frac{Q^2}{r_{i}^4}-\frac{\Lambda}{3}r_{i}^2\right)<0$ 
and $\left(1-\frac{5}{3}\frac{Q^2}{r_{i}^4}+\frac{\Lambda}{3}r_{i}^2\right)>0$ or 
$\left(1-\frac{1}{3}\frac{Q^2}{r_{i}^4}-\frac{\Lambda}{3}r_{i}^2\right)>0$ and 
$\left(1-\frac{5}{3}\frac{Q^2}{r_{i}^4}+\frac{\Lambda}{3}r_{i}^2\right)<0$, in this case 
the BH is thermodynamically stable.

Case II: The  specific  heat $C_{i}$ is negative when $\left(1-\frac{1}{3}\frac{Q^2}{r_{i}^4}-\frac{\Lambda}{3}r_{i}^2\right)>0$ 
and $\left(1-\frac{5}{3}\frac{Q^2}{r_{i}^4}+\frac{\Lambda}{3}r_{i}^2\right)>0$ or 
$\left(1-\frac{1}{3}\frac{Q^2}{r_{i}^4}-\frac{\Lambda}{3}r_{i}^2\right)<0$ and 
$\left(1-\frac{5}{3}\frac{Q^2}{r_{i}^4}+\frac{\Lambda}{3}r_{i}^2\right)<0$, in this case 
the BH is thermodynamically unstable.

Case III:  The  specific  heat $C_{i}$ blows up when $\left(1-\frac{5}{3}\frac{Q^2}{r_{i}^4}+\frac{\Lambda}{3}r_{i}^2\right)=0$, 
in this case the BH undergoes a second order phase transition.

\begin{figure}[h]
  \begin{center}
\subfigure[]{
\includegraphics[width=2.1in,angle=0]{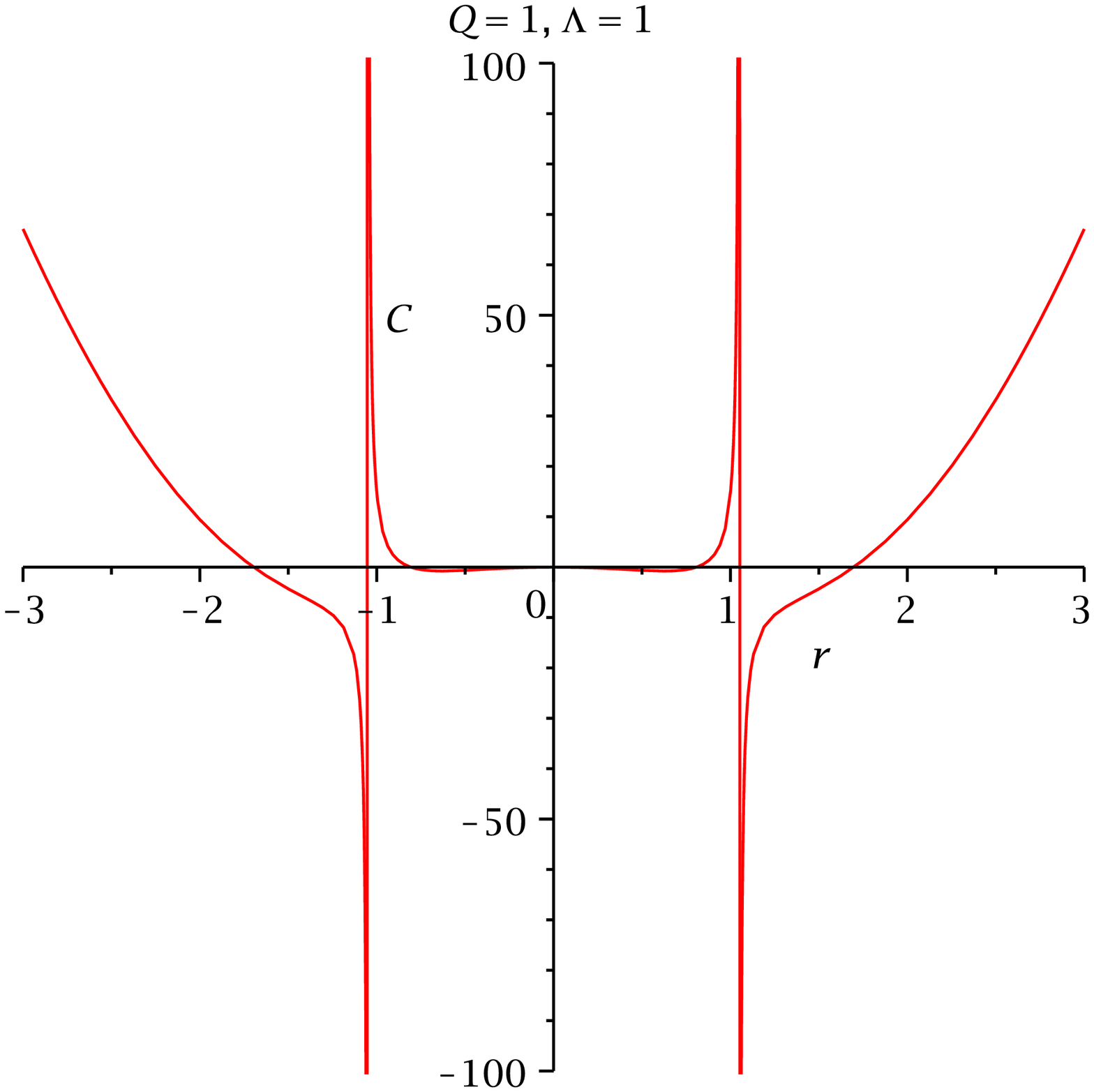}} 
\subfigure[]{
 \includegraphics[width=2.1in,angle=0]{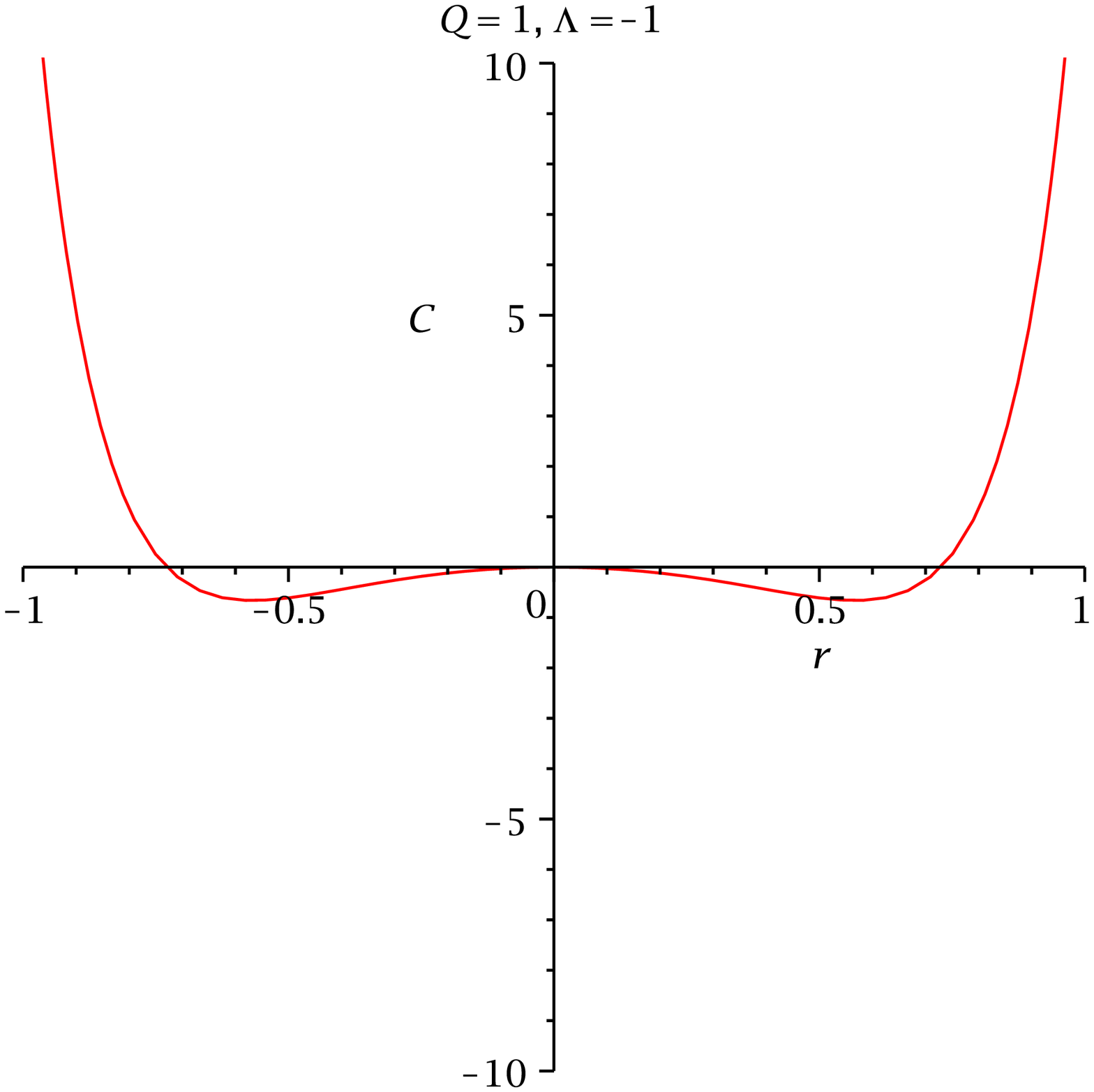}}
 \subfigure[]{
 \includegraphics[width=2.1in,angle=0]{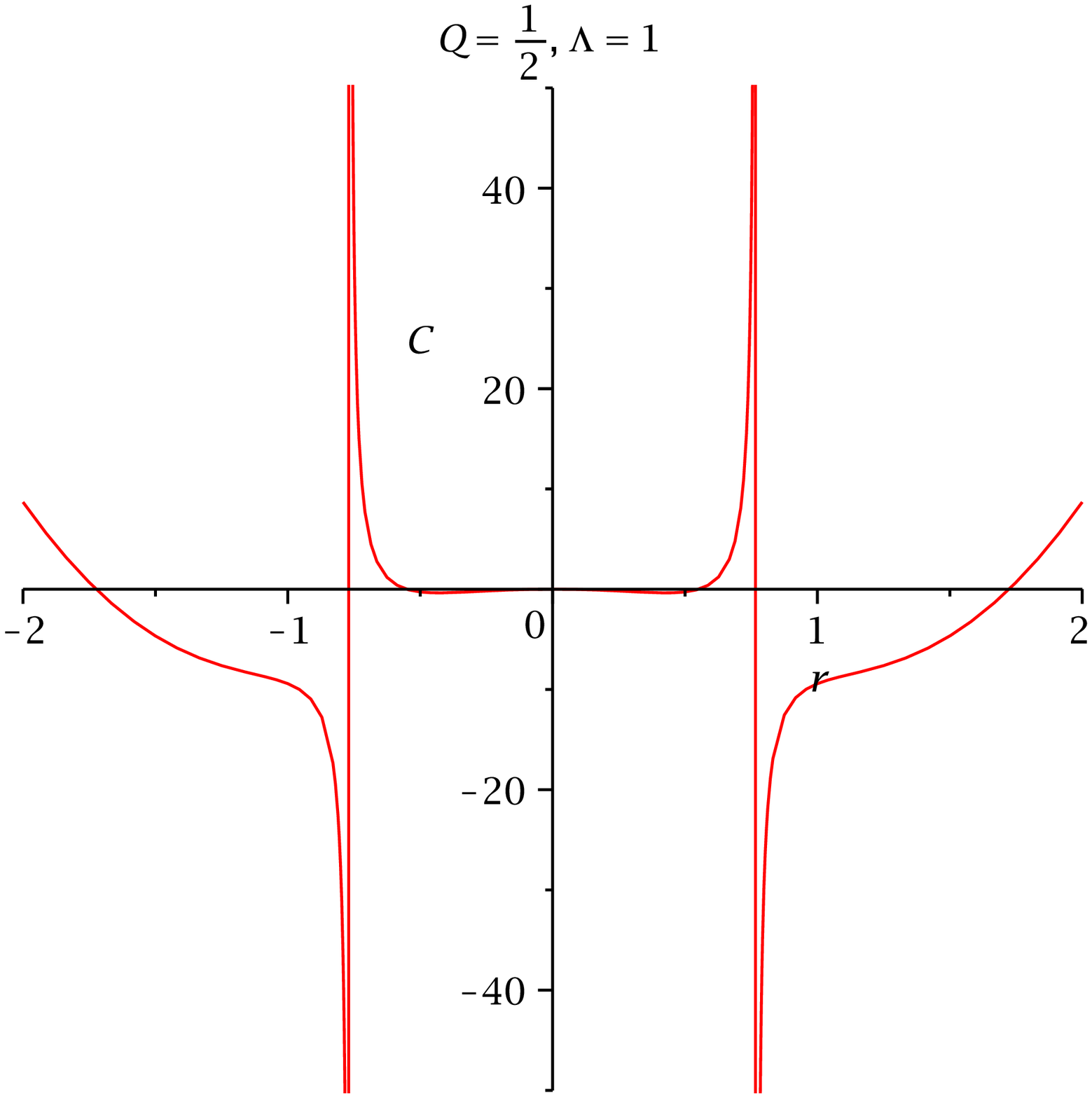}}
 \subfigure[ ]{
 \includegraphics[width=2.1in,angle=0]{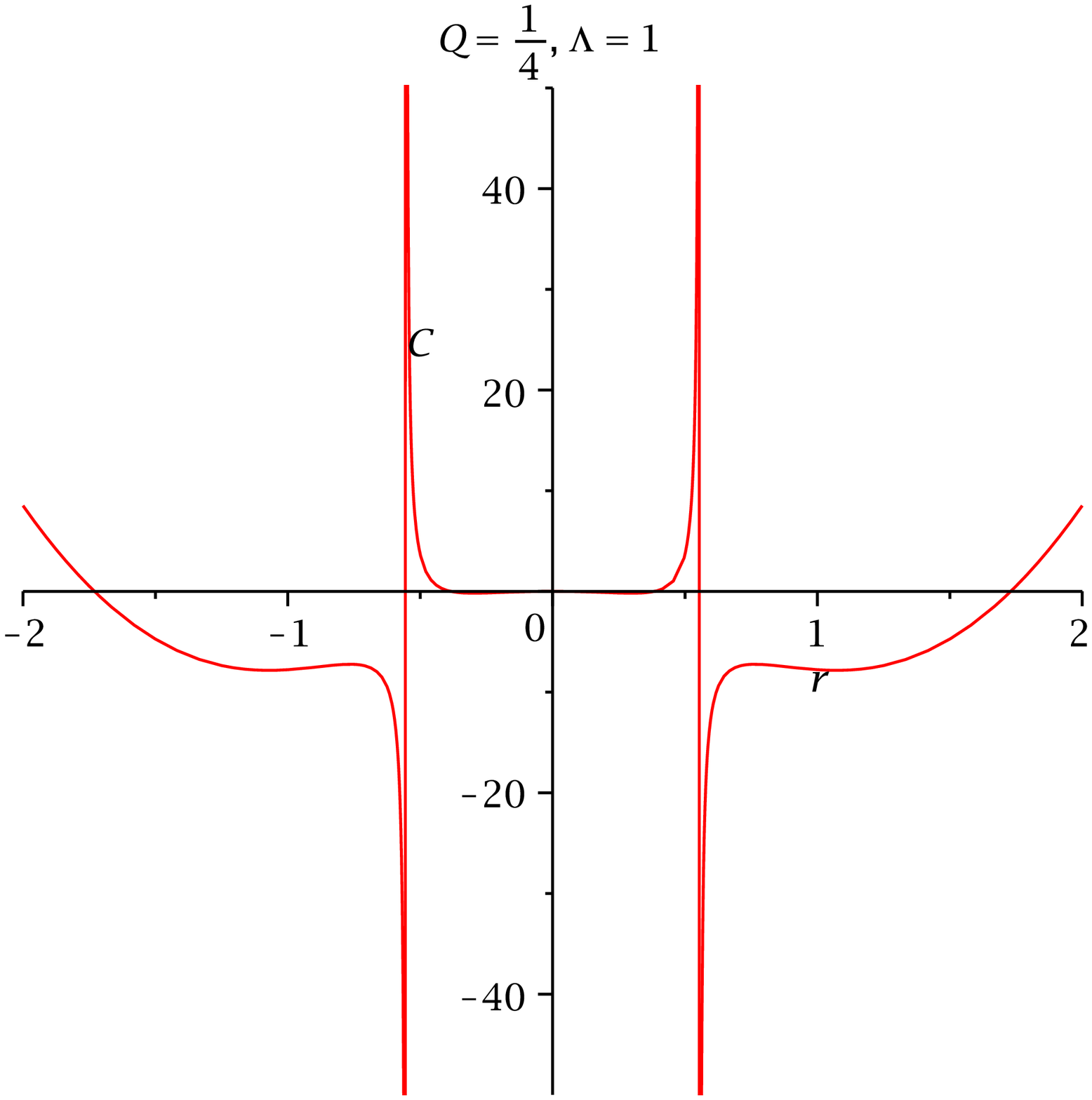}}
 \subfigure[]{
 \includegraphics[width=2.1in,angle=0]{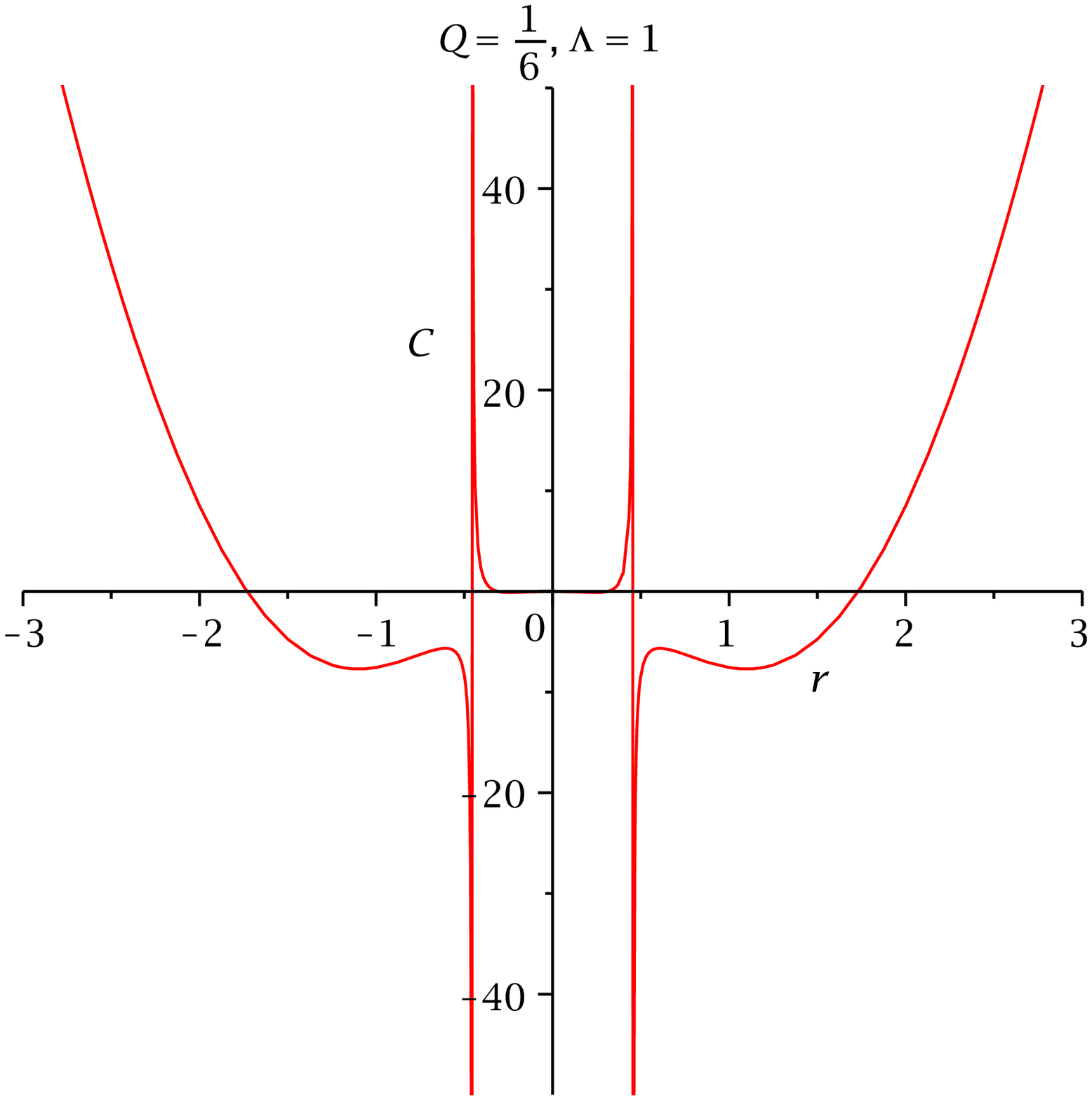}}
 \subfigure[]{
 \includegraphics[width=2.1in,angle=0]{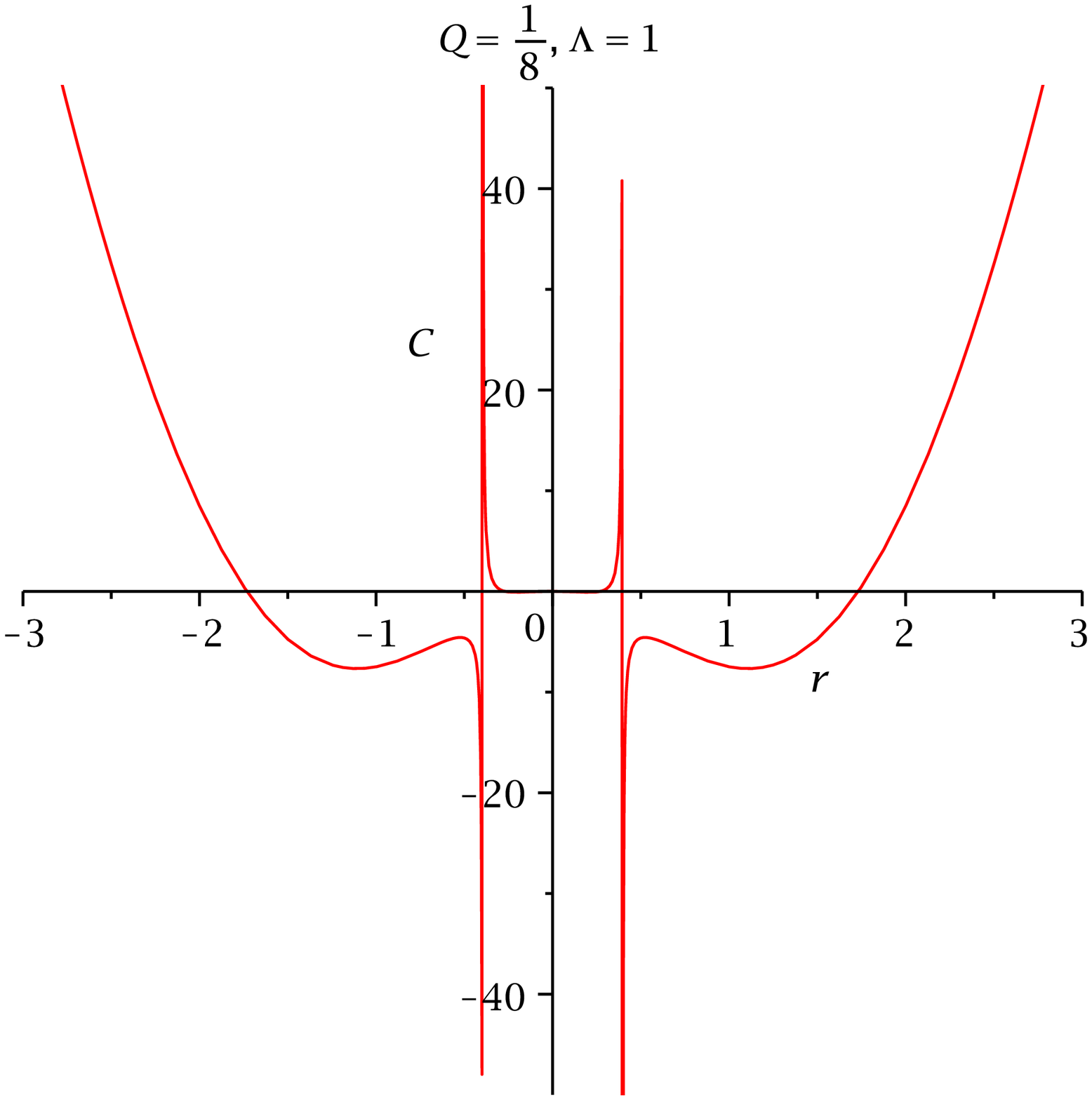}}
\caption{\label{fg1}\textit{ The figure shows the variation  of $C_{1}$  with $r_{1}$ for various values of $Q$ and $\Lambda$.}}
\end{center}
\end{figure}

\begin{figure}[h]
  \begin{center}
\subfigure[]{
\includegraphics[width=2.1in,angle=0]{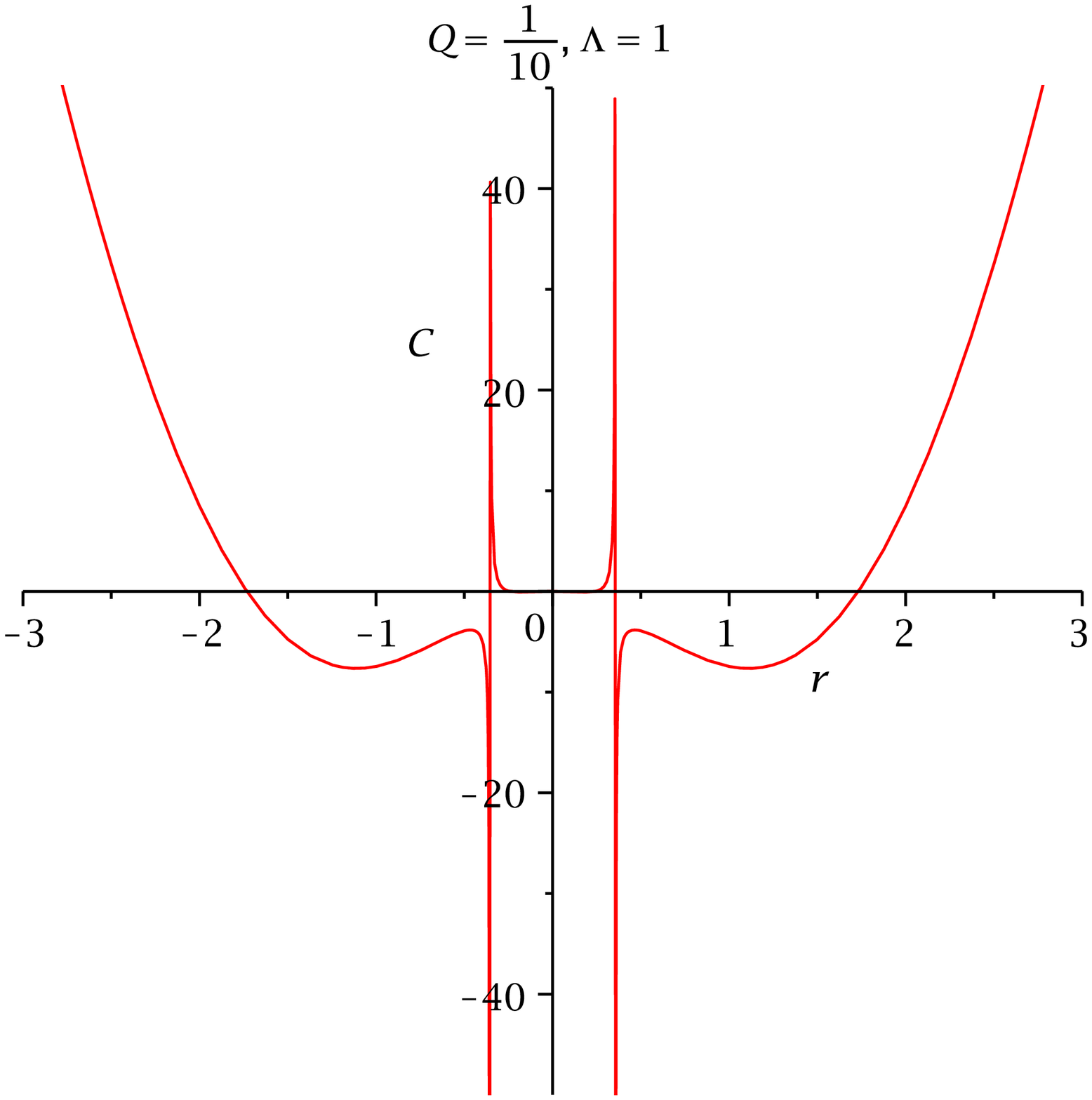}} 
\subfigure[]{
 \includegraphics[width=2.1in,angle=0]{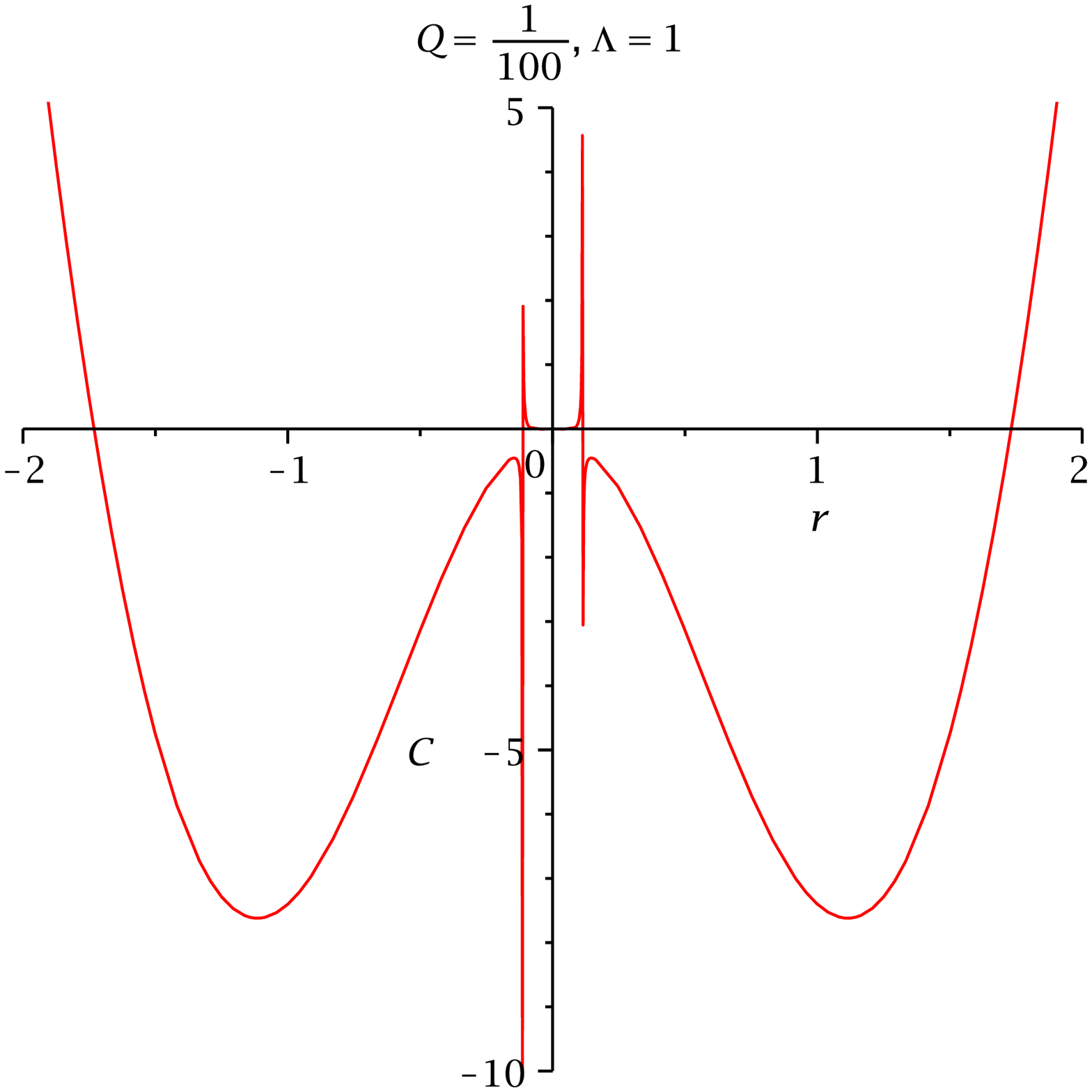}}
 \subfigure[]{
 \includegraphics[width=2.1in,angle=0]{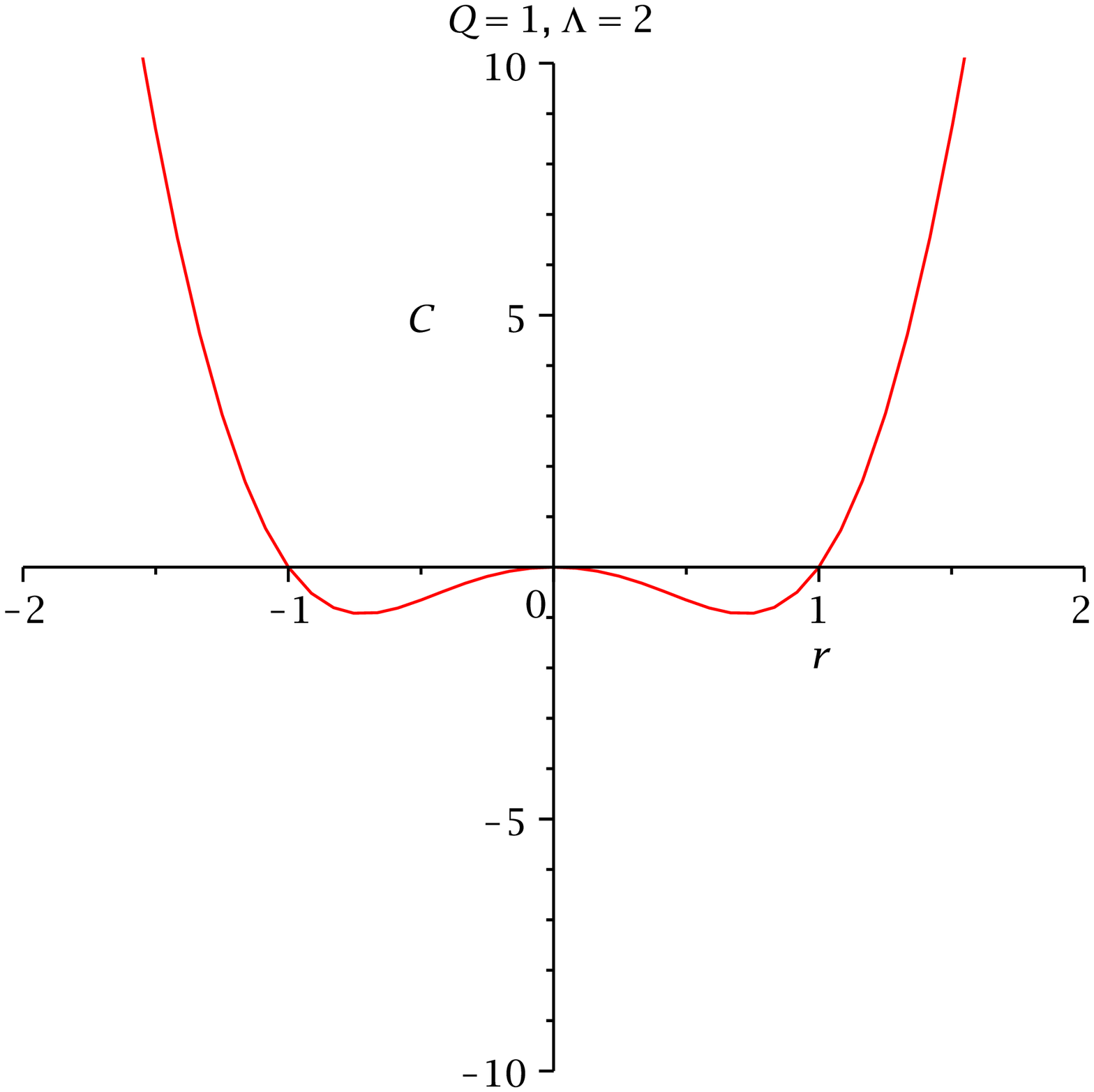}}
 \subfigure[ ]{
 \includegraphics[width=2.1in,angle=0]{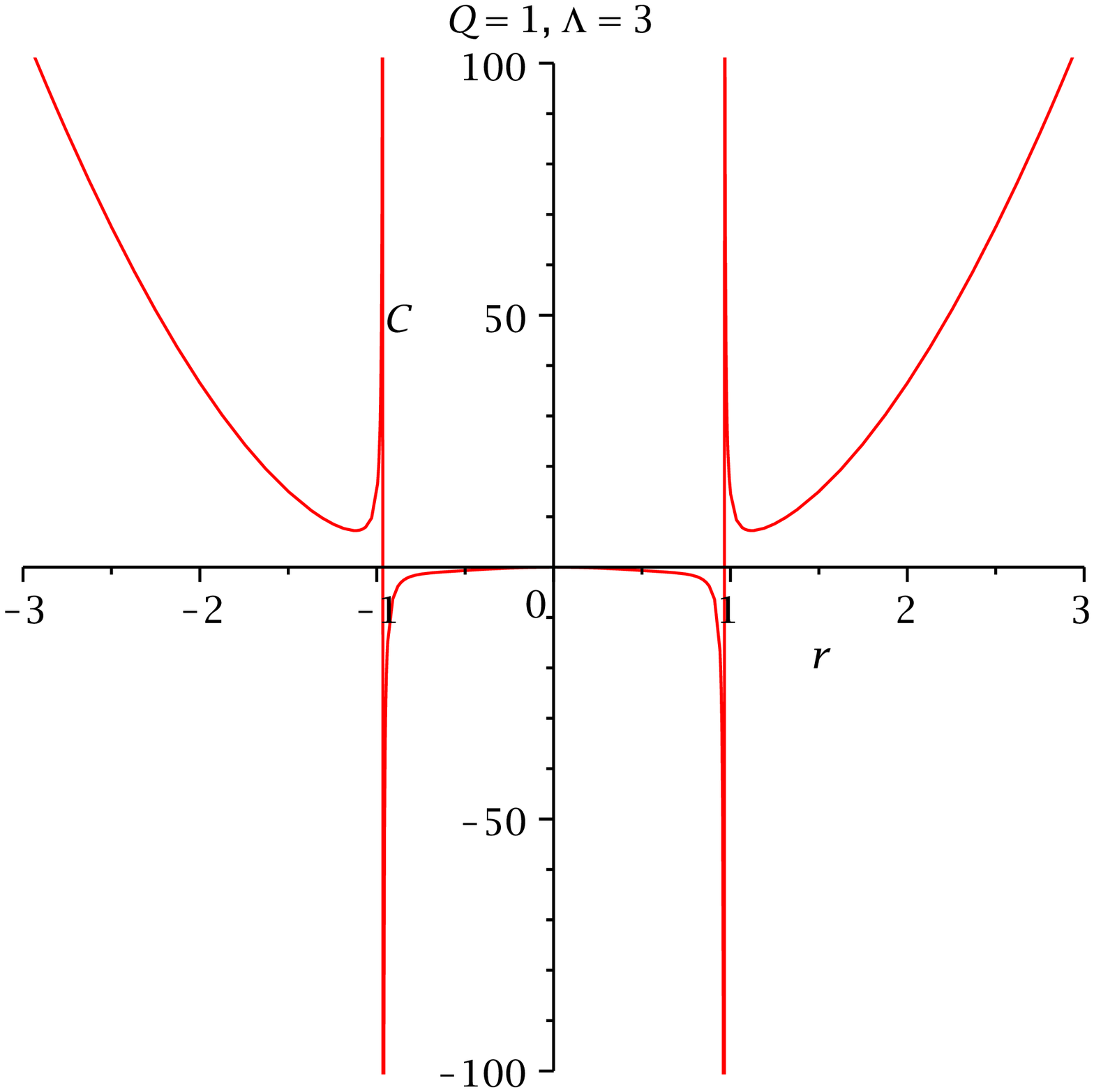}}
 \subfigure[]{
 \includegraphics[width=2.1in,angle=0]{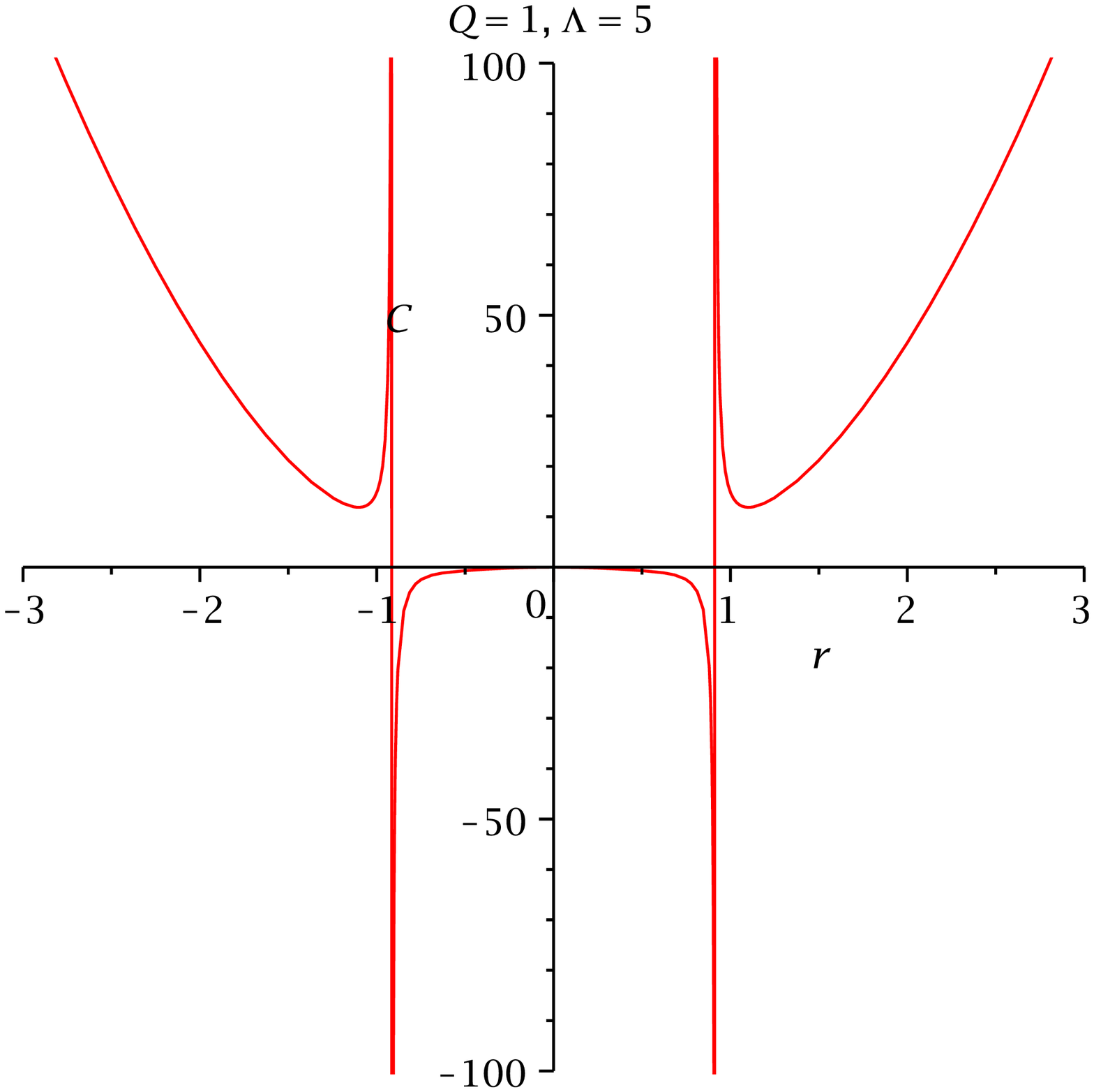}}
 \subfigure[]{
 \includegraphics[width=2.1in,angle=0]{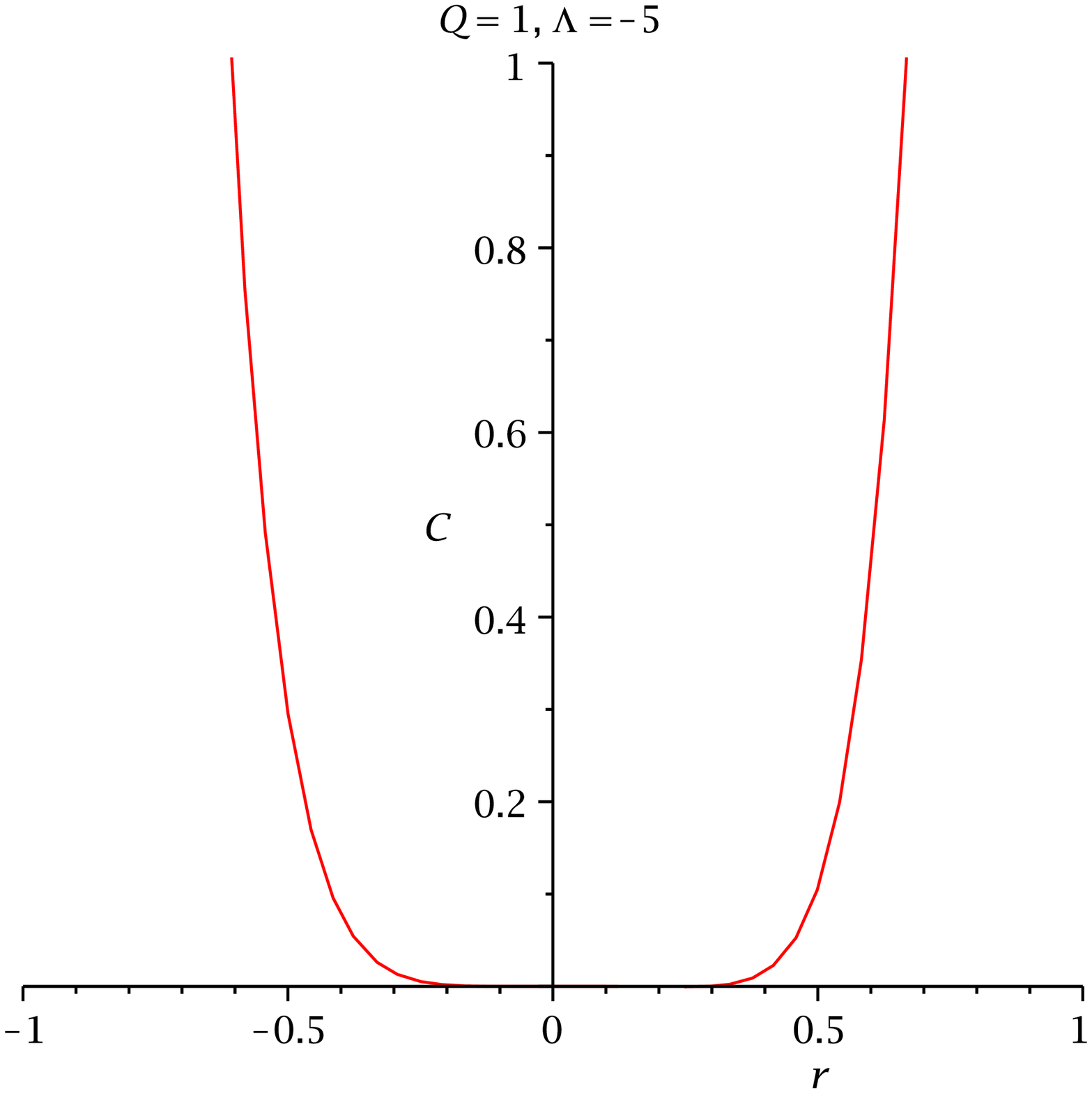}}
\caption{\label{fg2}\textit{ The figure shows the variation  of $C_{1}$  with $r_{1}$.}}
\end{center}
\end{figure}

It should be noted that when the Cosmological parameter goes to zero value  we obtain the result for EMGB BH.
In Fig. \ref{fg1} and Fig. \ref{fg2}, we have plotted the specific heat with the radial coordinate for various values 
of charge and cosmological constant. Form the diagram, it follows that how the specific heat changes sign from stability 
to instability region.

\section{\label{dis} Conclusion:}
In this work, we have derived area (or entropy) functional relation   
for 5D-Gauss-Bonnet-AdS Black Hole. We have found that some complicated function of three or two horizons area 
is indeed mass-independent. This could turn out to be  an ``universal'' quantity.
On the other hand, the curious result that we have obtained that the area product relation is 
mass-independent, whereas the entropy product relation is not mass-independent. 
This is mainly due to the fact that for higher curvature gravity the Bekenstein-Hawking entropy 
relation ${\cal S}=\frac{{\cal A}}{4}$ does not hold and it should be modified by some extra factor multiplied 
by the coupling constant.
When the coupling parameter goes to zero value we get the usual Bekenstein-Hawking entropy relation
as in Einstein gravity.  We also examined the local thermodynamic stability by computing the specific heat
of the BH. The BH posessess second order phase transition under appropriate condition. The first law 
of thermodynamics and Smarr-Gibbs-Duhem relation both are satisfied for this BH. In conclusion, the 
mass-independent area functional relation that we derived could turn to be an \emph{universal} quantity 
and gives us further understanding the nature of BH entropy (both inner and outer) at the microscopic 
level. In the \emph{Appendix} section we have derived the thermodynamic properties of  5D-EMGB BH with 
vanishing cosmological constant.

\section{Appendix}
In the appendix section we would like to compute the thermodynamic products of the 5D-EMGB BH without  
Cosmological constant. Also we shall derive the Smarr mass formula and Smarr-Gibbs-Duhem 
relation for this BH. The metric function for 5D static spherically metric without $\Lambda$ \cite{dehmami,taz} 
is given by
\begin{eqnarray}
{\cal F}(r) &=& k+\frac{r^2}{4\alpha}\pm\frac{r^2}{4\alpha}
\sqrt{1+\frac{8\alpha (M+2\alpha |k|)}{r^4}-\frac{8\alpha Q^2}{3r^6}} 
~.\label{app1}
\end{eqnarray}
where the geometric mass $M+2\alpha |k|$ is quite different from that of ADM mass $M$ of Einstein gravity for $k=\pm 1$. This 
is a special feature of the 5D EMGB but it has not been shown in higher dimensional Gauss-Bonnet gravity \cite{dehmami}.
It should be noted that this new mass is \emph{ADM mass plus $2|k|$ times of coupling constant}. It is also interesting to note that 
this mass contains \emph{coupling constant}. It is worthwhile to mention that for $\alpha>0$, the GB term increases the mass of the 
spacetime and for $\alpha<0$, the GB term decreases the mass of the spacetime. This is a special class of feature 
in 5D spacetime \cite{dehmami}.

The function ${\cal F}(r)$ is well defined if the expression 
$1+\frac{8\alpha (M+2\alpha |k|)}{r^4}-\frac{8\alpha Q^2}{3r^6}>0$ 
i.e. positive definite. The BH horizons exist when the condition ${\cal F}(r)=0$ is satisfied. The Killing horizon 
equation is given by 
\begin{eqnarray}
3kr_{i}^4- 3\left[(M+2\alpha |k|)-2\alpha k^2 \right]r_{i}^2 +Q^2  &=& 0 ~.\label{app2}
\end{eqnarray}
The solution of the equation is given by 
\begin{eqnarray}
r_{i} &=& \sqrt{\frac{\left(3M+6\alpha |k|-6\alpha k^2 \right)\pm \sqrt{\left(3M+6\alpha |k|-6\alpha k^2 \right)^2-12k Q^2}}{6k} 
} ~.\label{app3}
\end{eqnarray}
where $i=1,2$. 

For the special case $k=+1$ \footnote{For $\alpha<0$ and $k=+1$, we obtain a solution which is described asymtotically 
de-Sitter spacetime which represents a BH solution with an event horizon of radius $r_{1}$ provided that $3M^2 \geq 4Q^2$. 
More discussion regarding the $\pm$ branch solution of ${\cal F}(r)$ could be found in \cite{dehmami}.}  
(For $k=-1$, $r_{i}$ is imaginary thus it is unphysical)., we obtain the horizon radii for 5D-EMGB BH:
\begin{eqnarray}
r_{1} &=& \frac{1}{2} \left[\sqrt{M+\frac{2Q}{\sqrt{3}}} + \sqrt{M-\frac{2Q}{\sqrt{3}}}\right] \\ 
r_{2} &=& \frac{1}{2} \left[\sqrt{M+\frac{2Q}{\sqrt{3}}} -\sqrt{M-\frac{2Q}{\sqrt{3}}}\right]
~.\label{hrd}
\end{eqnarray}
Surprisingly, the horizon radii is independent of $\alpha$. It is due to the topological in nature.

The surface area \cite{ar} ${\cal H}^{1}$  and ${\cal H}^{2}$  is 
\begin{eqnarray}
{\cal A}_{1} &=& 2\pi^2 r_{1}^{3} \, \mbox{and}\, {\cal A}_{2} = 2\pi^2 r_{2}^{3}~.\label{ha}
\end{eqnarray}
and their product is 
\begin{eqnarray}
{\cal A}_{1}{\cal A}_{2}  &=& \frac{4}{3\sqrt{3}} \pi^4 Q^3
 ~.\label{ha1}
\end{eqnarray}
The mass of the BH can be expressed in terms of the area of both ${\cal H}^{1}$ and ${\cal H}^{2}$:
\begin{eqnarray}
M &=& 2\left(\frac{{\cal A}_{1}}{2\pi^2}\right)^{\frac{2}{3}}+
\frac{2Q^2}{3}\left(\frac{2\pi^2}{{\cal A}_{1}}\right)^{\frac{2}{3}} \\
M &=& 2\left(\frac{{\cal A}_{2}}{2\pi^2}\right)^{\frac{2}{3}}+
\frac{2Q^2}{3}\left(\frac{2\pi^2}{{\cal A}_{2}}\right)^{\frac{2}{3}} ~.\label{ha2}
\end{eqnarray}
The mass differential of ${\cal H}^{1}$ and ${\cal H}^{2}$ becomes
\begin{eqnarray}
dM &=& \Upsilon_{1} d{\cal A}_{1}  +\Phi_{1} dQ \\
dM &=& \Upsilon_{2} d{\cal A}_{2}  +\Phi_{2} dQ
~. \label{ap1}
\end{eqnarray}
where
\begin{eqnarray}
\Upsilon_{1} &=& \frac{\partial M}{\partial {\cal A}_{1}}= \frac{2}{3(2\pi^2)^{2/3}} 
\frac{1}{{\cal A}_{1}^{1/3}}\left[1-\frac{(2\pi^2)^{4/3}}{3}\frac{Q^2}{{\cal A}_{1}^{4/3}}\right] \\
\Upsilon_{2} &=& \frac{\partial M}{\partial {\cal A}_{2}}= \frac{2}{3(2\pi^2)^{2/3}} 
\frac{1}{{\cal A}_{2}^{1/3}}\left[1-\frac{(2\pi^2)^{4/3}}{3}\frac{Q^2}{{\cal A}_{2}^{4/3}}\right] \\
\Phi_{1} &=& \frac{\partial M}{\partial Q}=\frac{4Q}{3}
\left(\frac{2\pi^2}{{\cal A}_{1}}\right)^{\frac{2}{3}}\\
\Phi_{2} &=& \frac{\partial M}{\partial Q}=\frac{4Q}{3}
\left(\frac{2\pi^2}{{\cal A}_{2}}\right)^{\frac{2}{3}}
~. \label{ap2}
\end{eqnarray}
Thus the 1st law of thermodynamics of ${\cal H}^{1}$ and ${\cal H}^{2}$ reads
\begin{eqnarray}
dM &=& + \frac{T_{1}}{4} d{\cal A}_{1}  +\Phi_{1} dQ\\
dM &=& - \frac{T_{2}}{4} d{\cal A}_{2}  +\Phi_{2} dQ
~. \label{ap3}
\end{eqnarray}
where the BH temperature of ${\cal H}^{1}$ and ${\cal H}^{2}$ defined as
\begin{eqnarray}
T_{1} &=& \frac{16}{3(2\pi^2)^{2/3}} 
\frac{1}{{\cal A}_{1}^{1/3}}\left[1-\frac{(2\pi^2)^{4/3}}{3}\frac{Q^2}{{\cal A}_{1}^{4/3}} \right] \\
T_{2} &=& \frac{16}{3(2\pi^2)^{2/3}} 
\frac{1}{{\cal A}_{2}^{1/3}}\left[1-\frac{(2\pi^2)^{4/3}}{3}\frac{Q^2}{{\cal A}_{2}^{4/3}} \right]
~. \label{ap4}
\end{eqnarray}
The Smarr-Gibbs-Duhem relation of ${\cal H}^{1}$ and ${\cal H}^{2}$ becomes
\begin{eqnarray}
M &=& + \frac{T_{1}}{4} {\cal A}_{1}  +\Phi_{1} Q \\
M &=& - \frac{T_{2}}{4} {\cal A}_{2}  +\Phi_{2} Q ~. \label{ap0}
\end{eqnarray}

The specific heat of both ${\cal H}^{1}$ and ${\cal H}^{2}$ is 
\begin{eqnarray}
C_{1} &=& -\frac{3(2\pi^2)^{4/3}}{2} r_{1}^3 
\frac{\left(1-\frac{1}{3}\frac{Q^2}{r_{1}^4}\right)}{\left(1-\frac{5}{3}\frac{Q^2}{r_{1}^4}\right)} \\
C_{2} &=& -\frac{3(2\pi^2)^{4/3}}{2} r_{2}^3 \frac{\left(1-\frac{1}{3}\frac{Q^2}{r_{2}^4}\right)}
{\left(1-\frac{5}{3}\frac{Q^2}{r_{2}^4}\right)}
.~\label{ap5}
\end{eqnarray}
The phase transition point occurs  at $r_{1}=\left(\frac{5Q^2}{3}\right)^{1/4}$ and $r_{2}=\left(\frac{5Q^2}{3}\right)^{1/4}$.

Finally the Komar energy  of ${\cal H}^{1}$ and ${\cal H}^{2}$ is given by 
\begin{eqnarray}
E_{1} &=& \frac{8}{3(2\pi^2)^{1/3}} {r}_{1}^{2}\left[1-\frac{Q^2}{3{r}_{1}^{4}} \right] \\
E_{2} &=& \frac{8}{3(2\pi^2)^{1/3}} {r}_{2}^{2}\left[1-\frac{Q^2}{3{r}_{2}^{4}} \right]
~. \label{ap6}
\end{eqnarray}
From the above analysis, we can conclude that except the area product of ${\cal H}^{1}$ and ${\cal H}^{2}$, 
all other products of ${\cal H}^{1}$ and ${\cal H}^{2}$ are mass depended. 

\section*{Acknowledgements}
I am grateful to the anonyomous referee for several useful suggestions.


\end{document}